# Dynamic State Estimation for Integrated Natural Gas and Electric Power Systems

Liang Chen
School of Automation
Nanjing University of Information
Science and Technology
Nanjing, China
ch.lg@nuist.edu.cn

Xinxin Hui
Weifang Power Supply Company
State Grid Shandong Electric Power
Company
Weifang, China
1578716468@qq.com

Songlin Gu
State Grid Economic and
Technological Research Institute
CO.,LTD.
Beijing, China
gsl0516@163.com

Manyun Huang
College of Energy and Electrical
Engineering
Hohai University
Nanjing, China
hmy_hhu@yeah.net

Yang Li
School of Electrical Engineering
Northeast Electric Power University
Jilin, China
liyang@neepu.edu.cn

*Abstract*—a dynamic state estimation method of integrated natural gas and electric power systems (IGESs) in proposed. Firstly, the coupling model of gas pipeline networks and power systems by gas turbine units (GTUs) is established. Secondly, the Kalman filter based linear DSE model for the IGES is built. The gas density and mass flow rate, as well as the real and imaginary parts of bus voltages are taken as states, which are predicted by the linearized fluid dynamic equations of gases and exponential smoothing techniques. Boundary conditions of pipeline networks are used as supplementary constraints in the system model. At last, the proposed method is applied to an IGES including a 30-node pipeline network and IEEE 39-bus system coupled by two GTUs. Two indexes are used to evaluate the DSE performance under three measurement error conditions, and the results show that the DSE can obtain the accurate dynamic states in different conditions.

I. INTRODUCTION

The penetration of large-scale renewable energies increases the random fluctuation of power supplies, making the power system more vulnerable [1]. However, the gas turbine units (GTUs) can respond to the power fluctuation of renewable energies rapidly, which enhance the power system flexibility [2]. The integrated gas and electric power systems (IGESs) are becoming the hot research field [3].

The accurate states of IGESs are essential for the advanced applications. Although some important points are equipped with measurements, it is impossible to capture the entire useful states. On the other hand, the measuring devices experience random errors and bad data inevitably [4], [5], the data acquired from which cannot be applied directly before the state estimation.

The static state estimation based on weighted least squares (WLS) has been widely applied in power systems[6]. Practically, the dynamic states are more important for the system operation and control. To capture the power dynamics, the quasi static state estimation considering the slow changes of power loads have been proposed [7]. The Holt's exponential smoothing techniques [8] are used to predict the

This work is supported by the Natural Science Foundation of Jilin Province, China under Grant No. 2020122349JC.

loads, and the quasi static states are estimated by Kalman filter technology [9]. For the natural gas pipeline networks, the dynamic state is of great important for the optimal control and economical operating. The dynamic processes of pressures and mass flow rates of natural gas in the pipelines caused by the continuous changes of gas loads and supplies can be represented by a set of nonlinear partial differential equations (PDEs) being derived from the momentum conservation principles and materiel balances [10],[11]. Based on the PDEs, the state estimation for gas pipeline networks are proposed [12]-[14]. In [12] the extended Kalman filter (EKF) is applied to solve the state estimation problem. In [13], the nonlinear PDEs are linearized, and a pair of Kalman filter-based estimators running in parallel is carried out. In [14], a joint state and parameter estimation method for large scale networks of pipelines is presented based on the gradient descent algorithm. The study objects in these research works are decoupled with each other. In [15], a decentralized state estimation considering the spatiotemporal quasi dynamics of the heating pipeline networks for a whole combined heat and power system is proposed. Although many research works about state estimation problems have been came out, the research on the DSE for the IGESs is still a blank field.

To capture the accurate dynamic states, a DSE method for IGESs is put forward for the first time in this paper along with the following contributions:

1) The linear DSE model of the IGESs is established for the first time. An energy conversion coefficient is used to model the energy transformation between natural gases and electric powers through GTUs, based on which the coupling model of the gas pipeline networks and power systems by GTUs is established.

2) The Holt's exponential smoothing techniques are used to predict the power system states, and the linear power system dynamic equations are derived.

3) The PDEs of the gas pipelines are less than the states, so the PDEs cannot be solved. Aiming at this problem, boundary conditions are taken as supplementary constraints of the PDEs. However, the boundary conditions of the mass flow rate balance at pipeline nodes are algebraic equations, which are not able to predict the states. To solve this problem, the gas loads are also predicted by the Holt's exponential

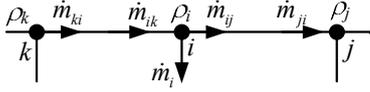

Fig. 1. The model of natural gas pipeline networks.

smoothing techniques, which are taken as control variables for the first time.

## II. MODEL OF POWER SYSTEM

The states can be predicted by the Holt's exponential smoothing techniques presented as:

$$L_t = \alpha_S x_{Et} + (1-\alpha_S)(L_{t-1}+T_{t-1}) \quad (1a)$$
$$T_t = \beta_S(L_t - L_{t-1}) + (1-\beta_S)T_{t-1} \quad (1b)$$
$$x_{Et+1} = L_t + T_t \quad (1c)$$

where, $x_{Et} = [e_{1t}\ f_{1t}\ e_{2t}\ f_{2t}\ \cdots\ e_{n_B t}\ f_{n_B t}]^T$ is the power system states, $x_{Et} \in \mathbb{R}^{2n_B \times 1}$, $n_B$ is the number of buses; $e_{it}$ and $f_{it}$ are the bus voltage real and imaginary parts at time $t$ respectively; $0<\alpha_S, \beta_S<1$ are the smoothing parameters. The initial values of $L_t$ and $T_t$ are $x_{E2}$ and $x_{E2} - x_{E1}$ respectively. Equation (1) can be written as the following matrix form:

$$x_{Et+1} = \alpha_S x_{Et} + u_{Et+1} \quad (2)$$

where, $u_{Et+1} = (1-\alpha_S)(L_{t-1}+T_{t-1}) + T_t$.

The measurement vector of power system at time instant $t$ is $z_{Et} = [\ z_{Vt}^T\ z_{IBt}^T\ z_{INt}^T\ ]^T$, in which $z_{Vt} = [e_{Z1t}\ f_{Z1t}\ e_{Z2t}\ f_{Z2t}\ \cdots\ e_{Zn_Bt}\ f_{Zn_Bt}]^T$, $z_{Vt} \in \mathbb{R}^{2n_B \times 1}$, $z_{IBt} = [\ \cdots\ I_{BRij,t}\ I_{BIij,t}\ I_{BRji,t}\ I_{BIji,t}\ \cdots\ ]^T$, $z_{IBt} \in \mathbb{R}^{4n_L \times 1}$, $z_{INt} = [I_{IR1t}\ I_{II1t}\ I_{IR2t}\ I_{II2t}\ \cdots\ I_{IRn_Bt}\ I_{IIn_Bt}]^T$, $z_{INt} \in \mathbb{R}^{2n_B \times 1}$. $e_{Zit}$ and $f_{Zit}$ are the measurements of voltage real and imaginary parts at bus $i$ respectively; $I_{BRij,t}$ and $I_{BIij,t}$ are the current real and imaginary part measurements of branch $i$-$j$ respectively; $n_L$ is the number of branches; $I_{IRit}$ and $I_{IIit}$ are the injected current real and imaginary part measurements of bus $i$ respectively. The function between $z_{Et}$ and $x_{Et}$ is

$$z_{Et} = \begin{bmatrix} z_{Vt} \\ z_{IBt} \\ z_{INt} \end{bmatrix} = \begin{bmatrix} I_{2n_B \times 2n_B} \\ H_{IB} \\ H_{IN} \end{bmatrix} x_{Et} = H_E x_{Et} \quad (3)$$

where, $I_{2n_B \times 2n_B}$ is a $2n_B$ dimensional identity matrix; $H_{IB}$ and $H_{IN}$ are the measurement matrixes of branch currents and injected currents, respectively.

## III. MODEL OF GAS PIPELINE NETWORKS

### A. Partial Differential Equations of Gas Flow in Pipeline

The fluid dynamic process of compressible gas along horizontal pipelines can be presented as a set of nonlinear PDEs [10][14]. To linearize the PDEs, the absolute value of average gas velocity is used. Omitting the convective effect of the natural gas, he linear PDEs can be presented as:

$$\frac{\partial \rho}{\partial \tau} + \frac{\partial \dot{m}}{a \partial \varsigma} = 0 \quad (4a)$$

$$\frac{\partial \dot{m}}{a \partial \tau} + c^2 \frac{\partial \rho}{\partial \varsigma} + \frac{\gamma \dot{m} |\overline{v}_G|}{2da} = 0 \quad (4b)$$

where, $\rho$ and $\dot{m}$ are the gas density and mass flow rate, respectively; $|\overline{v}_G|$ is the absolute value of average gas velocity; $\tau$ and $\varsigma$ are the time and space along pipelines respectively; a, $d$ and $\gamma$ are the cross-section area of pipelines, the pipeline diameter and the friction factor between natural gases and pipeline inner surfaces, respectively; $c$ is the sound speed.

The model of pipeline networks is shown in Fig. 1. The node numbers $i$, $j$ and $k$ satisfy $k< i < j$, and the gas flow direction is from the small number node to the big one. The mass flow rate going out from the sink node $i$ is $\dot{m}_i$. The linear PDEs (4) are solved by the Euler finite difference technique:

$$\rho_{j,t+1} - \rho_{j,t} + \rho_{i,t+1} - \rho_{i,t}$$
$$+ \frac{\Delta t}{L_{Pij} a_{ij}} \left( \dot{m}_{ji,t+1} - \dot{m}_{ij,t+1} + \dot{m}_{ji,t} - \dot{m}_{ij,t} \right) = 0 \quad (5a)$$

$$\dot{m}_{ji,t+1} - \dot{m}_{ij,t+1} + \dot{m}_{ji,t} - \dot{m}_{ij,t}$$
$$+ \frac{a_{ij} \Delta t c^2}{L_{Pij}} \left( \rho_{j,t+1} - \rho_{i,t+1} + \rho_{j,t} - \rho_{i,t} \right) \quad (5b)$$
$$+ \frac{\gamma |\overline{v}_G| \Delta t}{4 d_{ij} a_{ij}} \left( \dot{m}_{ij,t+1} + \dot{m}_{ji,t+1} + \dot{m}_{ij,t} + \dot{m}_{ji,t} \right) = 0$$

where, the subscripts $i$, $ij$ and $t$ represent node $i$, pipeline $i$-$j$, and time instance $t$, respectively; $L_{Pij}$ is the length of pipeline $i$-$j$; $\Delta t$ is the time step; $\dot{m}_{ij,t}$ is the mass flow rate from $i$ to $j$ in pipeline $i$-$j$.

### B. Boundary Conditions

In this paper, the nodes are classified into two types: source nodes and sink nodes. The gas density is constant at source nodes, and the mass flow rate is balanced at sink nods. These two features are named boundary conditions, which are also the constraints of gas states, represented as follows:

$$\rho_{s,t} = \rho_{Cs}, \quad s\text{: source node} \quad (6a)$$
$$\sum_k \dot{m}_{ik,t} - \sum_j \dot{m}_{ij,t} = \dot{m}_{i,t}, \ k, j \in i, \ k<i, j>i, \ i\text{: sink node} \quad (6b)$$

where, $k, j \in i$ represents node $k$ and $j$ are connected to $i$; $\rho_{Cs}$ is the constant gas density at node $s$; $\dot{m}_{i,t}$ is the mass flow rate going out from node $i$ at time $t$.

### C. System Equation

The gas densities and mass flow rates at the two ends of pipelines are taken as states. For a $n_N$-node and $n_P$-pipeline gas network, the number of states is $n_N + 2n_P$. The state vector at time $t$ is $x_{Gt} = [x_{Gr,t},\ x_{Gm,t}]^T$, $x_{Gr,t} = [\rho_{1,t}, \rho_{2,t}, \cdots, \rho_{n_N,t}]^T$, $x_{Gm,t} = [\cdots, \dot{m}_{ij,t}, \dot{m}_{ji,t}, \cdots]^T$. $x_{Gr,t} \in \mathbb{R}^{n_N \times 1}$, $x_{Gm,t} \in \mathbb{R}^{2n_P \times 1}$. Equation (5) can be rewritten in the following matrix form:

$$\begin{bmatrix} A_{11} & A_{12} \\ A_{21} & A_{22} \end{bmatrix} \begin{bmatrix} x_{Gr,t+1} \\ x_{Gm,t+1} \end{bmatrix} = \begin{bmatrix} A_{11} & -A_{12} \\ -A_{21} & -A_{22} \end{bmatrix} \begin{bmatrix} x_{Gr,t} \\ x_{Gm,t} \end{bmatrix} \quad (7)$$

$$A_{11}(l,i) = \begin{cases} 1, & \text{if } i \propto l \\ 0, & \text{else} \end{cases}, A_{11} \in \mathbb{R}^{n_P \times n_N} \quad (8)$$

$$A_{12} = \begin{bmatrix} -\xi_1 & \xi_1 & & & 0 \\ & -\xi_2 & \xi_2 & & \\ & & \ddots & & \\ 0 & & & -\xi_{n_L} & \xi_{n_L} \end{bmatrix}, A_{12} \in \mathbb{R}^{n_P \times 2n_P} \quad (9)$$

$$\xi_l = \frac{\Delta t}{L_{ij} a_{ij}}, i \propto l, j \propto l \quad (10)$$

$$A_{21}(l,i) = \begin{cases} -\beta_l, & \text{if } i \propto l, j \propto l, i < j \\ \beta_l, & \text{if } i \propto l, j \propto l, i > j \\ 0, & \text{else} \end{cases}, A_{21} \in \mathbb{R}^{n_P \times n_N} \quad (11)$$

$$\beta_l = \frac{a_{ij} \Delta t c^2}{L_{ij}}, i \propto l, j \propto l \quad (12)$$

$$A_{22}(l,i) = \begin{cases} \gamma_l - 1, & \text{if } i \propto l, j \propto l, i < j \\ \gamma_l + 1, & \text{if } i \propto l, j \propto l, i > j \\ 0, & \text{else} \end{cases}, A_{22} \in \mathbb{R}^{n_P \times 2n_P} \quad (13)$$

$$\gamma_l = \frac{f |\overline{v}_G| \Delta t}{4 d_{ij} a_{ij}}, i \propto l, j \propto l \quad (14)$$

where, $i \propto l$ means node $i$ is one end of pipeline $l$. The matrix form of boundary conditions (6) is

$$\begin{bmatrix} B_{11} & 0_{n_S \times 2n_P} \\ 0_{n_{SI} \times n_N} & B_{22} \end{bmatrix} \begin{bmatrix} x_{r,t} \\ x_{m,t} \end{bmatrix} = \begin{bmatrix} u_{r,t} \\ u_{m,t} \end{bmatrix} \quad (15)$$

$$B_{11}(s,i) = \begin{cases} 1, & i: \text{source node} \\ 0, & i: \text{sink node} \end{cases}, B_{11} \in \mathbb{R}^{n_S \times n_N} \quad (16)$$

$$B_{22}(j, 2l) = \begin{cases} 1, & j: \text{sink node}, j \propto l \\ 0, & \text{else} \end{cases}, B_{22} \in \mathbb{R}^{n_{SI} \times 2n_P} \quad (17)$$

$$B_{22}(i, 2l-1) = \begin{cases} -1, & i: \text{sink node}, i \propto l \\ 0, & \text{else} \end{cases} \quad (18)$$

where, $u_{r,t} = [\ldots, \rho_{Cs}, \ldots]^T$, $s$ is source node, $u_{r,t} \in \mathbb{R}^{n_S}$; $u_{m,t} = [\ldots, \dot{m}_{i,t}, \ldots]^T$, $i$ is sink node, $u_{m,t} \in \mathbb{R}^{n_{SI}}$; $n_S$ and $n_{SI}$ are the numbers of source nodes and sink nodes respectively; $\mathbf{0}$ is the zero matrix, the subscript of which is the dimension.

Equation (7) and (15) can be written as

$$\mathcal{A} x_{Gt+1} = \mathcal{B} x_{Gt} + \mathcal{U}_{t+1} \quad (19)$$

$$\mathcal{A} = \begin{bmatrix} A_{11} & A_{12} \\ A_{21} & A_{22} \\ B_{11} & 0_{n_S \times 2n_P} \\ 0_{n_{SI} \times n_N} & B_{22} \end{bmatrix}, \mathcal{B} = \begin{bmatrix} A_{11} & -A_{12} \\ -A_{21} & -A_{22} \\ 0_{n_S \times n_N} & 0_{n_S \times 2n_P} \\ 0_{n_{SI} \times n_N} & 0_{n_{SI} \times 2n_P} \end{bmatrix}, \mathcal{U}_{t+1} = \begin{bmatrix} 0_{n_P \times 1} \\ 0_{n_P \times 1} \\ u_{r,t+1} \\ u_{m,t+1} \end{bmatrix}.$$

The general system model of pipeline networks is obtained

$$x_{Gt+1} = F_G x_{Gt} + u_{Gt+1} \quad (20)$$

where, $F_G = \mathcal{A}^{-1} \mathcal{B}$, $u_{Gt+1} = \mathcal{A}^{-1} \mathcal{U}_{t+1}$.

### D. Measurement Equation

Assume that the node pressures and mass flow rates can be measured. The measurement equation is

$$z_{Gt+1} = H_G x_{Gt+1} \quad (21)$$

$$H_G = \begin{bmatrix} c^2 I_{n_N \times n_N} & 0_{n_N \times 2n_P} \\ 0_{n_N \times n_N} & H' \end{bmatrix} \quad (22)$$

$$\begin{cases} H'(j, 2l) = 1 \\ H'(i, 2l-1) = -1 \end{cases} i, j \propto l, i < j, H' \in \mathbb{R}^{n_N \times 2n_P} \quad (23)$$

where, $z_{Gt+1} = [z_{r1,t+1}, z_{r2,t+1}, \ldots, z_{rn_N,t+1}, z_{m1,t+1}, z_{m2,t+1}, \ldots,$ $z_{mn_N,t+1}]^T$; $z_{ri,t+1}$ and $z_{mi,t+1}$ are the pressure and mass flow rate measurements at time $t+1$ respectively.

## IV. DYNAMIC STATE ESTIMATION OF IGESs

### A. Coupling Model of IGESs

The relationship between the consumed gas and produced power of GTUs is represented as

$$P_{Gi,t} = \eta_i \dot{m}_{s,t}, i \prec s \quad (24)$$

where, $P_{Gi,t}$ and $\eta_i$ are the output power at time $t$ and energy conversion coefficient of the GTU connected to bus $i$; $i \prec s$ represents that the gas source of the GTU connected to bus $i$ is the sink node $s$. In power system, $P_{Gi,t}$ satisfy the bus injected power equation, which is shown as follows:

$$P_{Gi,t} = \sum_j [e_{it}(G_{Yij} e_{jt} - B_{Yij} f_{jt}) + f_{it}(G_{Yij} f_{jt} + B_{Yij} e_{jt})], j \mapsto i \quad (25)$$

where, $G_{Yij}$ and $B_{Yij}$ are the real and imaginary parts of the $i$th row $j$th column element in power system admittance matrix, respectively; $j \mapsto i$ represents bus $j$ is connected to bus $i$. The relationship between $P_{Gi,t}$ and power system states can be derived from (24), (25):

$$\dot{m}_{s,t} = \frac{1}{\eta_i} \sum_j [e_{it}(G_{Yij} e_{jt} - B_{Yij} f_{jt}) + f_{it}(G_{Yij} f_{jt} + B_{Yij} e_{jt})] \quad (26)$$

### B. Dynamic State Estimation Model of IGESs

The state vector of IGESs $x_{It}$ includes the electric power system states $x_{Et}$ and natural gas pipeline network states $x_{Gt}$, i.e., $x_{It} = [x_{Et}^T \ x_{Gt}^T]^T$, and the measurements $z_{It} = [z_{Et}^T \ z_{Gt}^T]^T$. The DSE model of IGESs is

$$\begin{cases} x_{It+1} = F_I x_{It} + u_{It+1} + v \\ z_{It+1} = H_I x_{It+1} + w \end{cases} \quad (27)$$

where, $F_I = \begin{bmatrix} \alpha_S I_{2n_B \times 2n_B} & 0_{2n_B \times (2n_P + n_N)} \\ 0_{2n_P \times 2n_B} & F_G \end{bmatrix}$, $u_{It+1} = [u_{Et+1}^T \ u_{Gt+1}^T]^T$,

$H_I = \begin{bmatrix} H_E & 0_{4(n_B + n_L) \times (2n_P + n_N)} \\ 0_{2n_N \times 2n_B} & H_G \end{bmatrix}$; $v$ and $w$ are the predicting

and measurement errors, respectively, $v \in \mathbb{R}^{2n_B + n_N + 2n_P}$, $w \in \mathbb{R}^{4n_B + 4n_L + 2n_N}$. The variance matrixes of $v$ and $w$ are $Q$ and $R$ respectively.

The value of $u_{It+1}$ can be given by the following method:
- For gas densities at source nodes, the values are constant;
- For the sink node that supplies gases to GTUs, the values are the mass flows at time $t+1$, obtained by (26). The voltage $e_{it+1}$ and $f_{it+1}$ in (26) are predicted by the Holt's exponential smoothing techniques.
- For the other sink nodes, the values are the mass flow rates of gas loads at time $t+1$, predicted by the Holt's exponential smoothing techniques directly.

Based on the model of (27), Kalman filter can be applied to solve the dynamic state estimation problem of IGESs.

## V. CASE STUDIES

To testify the validity of the proposed state estimator, an IGES with a 30-node natural gas pipeline network and IEEE39-bus testing system coupled by two GTUs is constructed, shown in Fig. 2. The dynamics of the electric power loads and GTUs' output power are simulated artificially. The power system states are obtained by power flow calculation using Matpower [16]. The mass flow rates of GTUs are calculated by (24) according to the output power. The dynamic states of pipeline networks are calculated by (20). The simulated values are true values, while the measured values are generated by adding specific distributed random numbers to the true values. The estimating step is 10 minutes, and the simulation time is 24 hours.

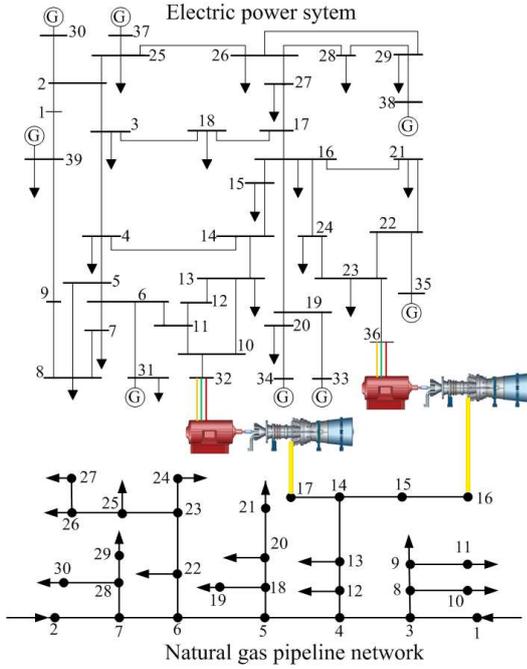

Fig. 2. The IGES testing system. The electric power system and natural gas pipeline network are coupled by two GTUs supplied natural gas by node 16 and 17 respectively. The bus 32 and 36 are powered by the GTUs.

The estimation is carried out under the following conditions:

a. White Gaussian noises. The measurement errors are white Gaussian noises.

b. Non-zero mean noises. The measurement errors obey normal distribution, but the means are not zero.

c. Non-Gaussian noises. The measurement errors are Non-Gaussian.

The value of GTU energy conversion coefficient here is 20.148 MW·s/kg under 40% efficiency. The lengths and cross-sectional diameters of the pipelines are given in Tab. I. The friction factor $\gamma$ and the gas speed $c^2$ are 0.015 and 340 m/s respectively. Pipeline node 1 and 2 are source nodes, the pressures of which are 41.48 bar and 41.63 bar respectively, being constant in the dynamic processes. The smoothing parameters $\alpha_S$ and $\beta_S$ are 0.8 and 0.7, respectively.

*A. White Gaussian Noises*

The DSE is carried out under the condition that all of the measurement errors are white Gaussian noises. The estimated states of the gas pipeline network and power system are shown in Fig. 3 and Fig. 4 respectively. The standard deviation of measurement errors is 0.02. It shows that the estimated value curve fits the true value curve much better than the measurements.

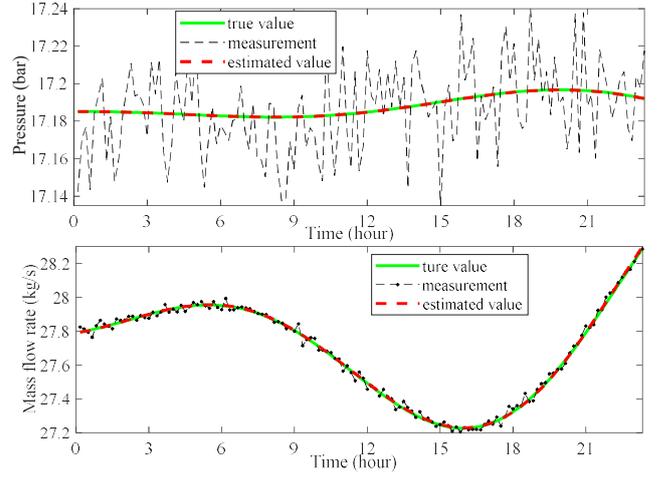

Fig. 3. Estimating results of the pressure and mass flow rate at pipeline node 16 with 0.02 measurement errors.

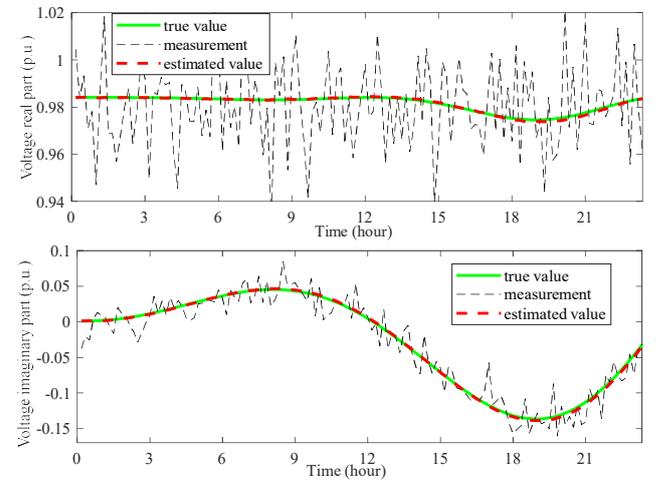

Fig. 4. Estimating results of the real and imaginary parts of voltage at bus 32 with 0.02 measurement errors.

To evaluate the performance of the DSE quantitatively, the filter coefficient $\varepsilon_1$ and the total variance of estimation errors $\varepsilon_2$ are defined [17], [18]:

$$\varepsilon_1 = \sum_{t=1}^{S} (\hat{x}_t - x_t^+)^2 \bigg/ \sum_{t=1}^{S} (x_t^M - x_t^+)^2 \qquad (28)$$

$$\varepsilon_2 = \sum_{t=1}^{S} (\hat{x}_t - x_t^+)^2 \bigg/ S \qquad (29)$$

where, $\hat{x}_t$, $x_t^M$ and $x_t^+$ are the estimated value, measured value and true value, respectively; $S$ is the sampling number.

The filter coefficients and total variances of gas pipeline states and electric power states are shown in Tab. I and Tab. II respectively. It can be seen that all of the filter coefficients are less than 1, meaning that the DSE method is effective. The total variances of estimation errors are less than the measurement error variance 0.0004, which means the estimated variances are less than the measurement variances.

*B. Non-zero Mean Noises*

In practical systems, the measurement errors do not always satisfy the zero-mean distribution. Based on the measurements in section *B*, the deviations ranged from 0.01 to 0.04 are added to the pressure measurement of node 14, mass flow rate measurement of node 21, voltage real and imaginary parts measurements of bus 11, to test the performances of the proposed DSE.

TABLE I. FILTER COEFFICIENTS AND TOTAL VARIANCES OF NODES

| Node | Pressure $\varepsilon_1(10^{-4})$ | Pressure $\varepsilon_2(10^{-8})$ | Mass flow rate $\varepsilon_1$ | Mass flow rate $\varepsilon_2(10^{-3})$ |
|---|---|---|---|---|
| 1 | / | / | 0.0374 | 0.0140 |
| 2 | / | / | 0.0396 | 0.0179 |
| 3 | 0.0004 | 0.0018 | 0.1806 | 0.0750 |
| 4 | 0.0001 | 0.0006 | 0.1804 | 0.0698 |
| 5 | 0.0007 | 0.0028 | 0.1818 | 0.0827 |
| 6 | 0.0019 | 0.0078 | 0.1837 | 0.0723 |
| 7 | 0.0028 | 0.0113 | 0.1835 | 0.0519 |
| 8 | 0.0008 | 0.0029 | 0.1815 | 0.0700 |
| 9 | 0.0191 | 0.0779 | 0.1103 | 0.0505 |
| 10 | 0.0193 | 0.0671 | 0.0394 | 0.0180 |
| 11 | 0.0166 | 0.0760 | 0.0398 | 0.0146 |
| 12 | 0.0025 | 0.0101 | 0.1080 | 0.0413 |
| 13 | 0.0030 | 0.0131 | 0.1075 | 0.0416 |
| 14 | 0.0060 | 0.0237 | 0.1741 | 0.0814 |
| 15 | 0.0109 | 0.0472 | 0.1103 | 0.0425 |
| 16 | 0.0115 | 0.0446 | 0.0694 | 0.0314 |
| 17 | 0.2022 | 0.9031 | 0.0495 | 0.0178 |
| 18 | 0.0066 | 0.0280 | 0.1830 | 0.0736 |
| 19 | 0.0201 | 0.0910 | 0.0398 | 0.0181 |
| 20 | 0.0059 | 0.0291 | 0.1106 | 0.0415 |
| 21 | 0.0458 | 0.1826 | 0.0398 | 0.0183 |
| 22 | 0.0132 | 0.0517 | 0.1111 | 0.0448 |
| 23 | 0.0079 | 0.0346 | 0.1795 | 0.0817 |
| 24 | 0.0108 | 0.0485 | 0.0394 | 0.0146 |
| 25 | 0.0098 | 0.0371 | 0.1064 | 0.0444 |
| 26 | 0.0282 | 0.0963 | 0.1111 | 0.0425 |
| 27 | 0.0116 | 0.0478 | 0.0396 | 0.0187 |
| 28 | 0.0069 | 0.0313 | 0.1809 | 0.0723 |
| 29 | 0.0205 | 0.0861 | 0.0399 | 0.0159 |
| 30 | 0.0120 | 0.0448 | 0.0395 | 0.0186 |

TABLE II. FILTER COEFFICIENTS AND TOTAL VARIANCES OF BUSES

| Bus | $e$ $\varepsilon_1$ | $e$ $\varepsilon_2(10^{-5})$ | $f$ $\varepsilon_1$ | $f$ $\varepsilon_2(10^{-5})$ |
|---|---|---|---|---|
| 1 | 0.0040 | 0.1618 | 0.0037 | 0.1548 |
| 2 | 0.0038 | 0.1529 | 0.0034 | 0.1493 |
| 3 | 0.0050 | 0.1499 | 0.0037 | 0.1488 |
| 4 | 0.0037 | 0.1484 | 0.0034 | 0.1462 |
| 5 | 0.0035 | 0.1480 | 0.0050 | 0.1483 |
| 6 | 0.0038 | 0.1479 | 0.0039 | 0.1485 |
| 7 | 0.0043 | 0.1490 | 0.0042 | 0.1479 |
| 8 | 0.0042 | 0.1486 | 0.0037 | 0.1484 |
| 9 | 0.0042 | 0.1680 | 0.0035 | 0.1577 |
| 10 | 0.0038 | 0.1504 | 0.0036 | 0.1485 |
| 11 | 0.0037 | 0.1491 | 0.0043 | 0.1484 |
| 12 | 0.0039 | 0.1646 | 0.0038 | 0.1623 |
| 13 | 0.0044 | 0.1500 | 0.0035 | 0.1492 |
| 14 | 0.0049 | 0.1491 | 0.0033 | 0.1462 |
| 15 | 0.0037 | 0.1510 | 0.0039 | 0.1515 |
| 16 | 0.0029 | 0.1511 | 0.0034 | 0.1491 |
| 17 | 0.0039 | 0.1530 | 0.0036 | 0.1476 |
| 18 | 0.0033 | 0.1512 | 0.0040 | 0.1487 |
| 19 | 0.0049 | 0.1519 | 0.0033 | 0.1420 |
| 20 | 0.0050 | 0.1566 | 0.0038 | 0.1436 |
| 21 | 0.0041 | 0.1579 | 0.0047 | 0.1535 |
| 22 | 0.0059 | 0.1635 | 0.0046 | 0.1618 |
| 23 | 0.0044 | 0.1631 | 0.0039 | 0.1594 |
| 24 | 0.0037 | 0.1510 | 0.0035 | 0.1499 |
| 25 | 0.0035 | 0.1519 | 0.0040 | 0.1536 |
| 26 | 0.0049 | 0.1564 | 0.0033 | 0.1491 |
| 27 | 0.0042 | 0.1574 | 0.0034 | 0.1485 |
| 28 | 0.0056 | 0.1781 | 0.0036 | 0.1629 |
| 29 | 0.0049 | 0.1778 | 0.0033 | 0.1616 |
| 30 | 0.0041 | 0.1602 | 0.0039 | 0.1485 |
| 31 | 0.0040 | 0.1465 | 0.0045 | 0.1518 |
| 32 | 0.0032 | 0.1605 | 0.0047 | 0.1535 |
| 33 | 0.0052 | 0.1496 | 0.0035 | 0.1403 |
| 34 | 0.0041 | 0.1635 | 0.0039 | 0.1464 |
| 35 | 0.0044 | 0.1731 | 0.0039 | 0.1664 |
| 36 | 0.0046 | 0.1720 | 0.0041 | 0.1616 |
| 37 | 0.0041 | 0.1586 | 0.0038 | 0.1597 |
| 38 | 0.0043 | 0.1783 | 0.0052 | 0.1691 |
| 39 | 0.0048 | 0.1728 | 0.0051 | 0.1618 |

TABLE III. THE VALUES OF INDEXES UNDER COLORED NOISES

| | | | Measurement error deviation | | | |
|---|---|---|---|---|---|---|
| | | | 0.01 | 0.02 | 0.03 | 0.04 |
| Node 14 | Pressure | $\varepsilon_1(10^{-5})$ | 1.67 | 3.20 | 3.53 | 4.22 |
| | | $\varepsilon_2(10^{-9})$ | 8.44 | 27.88 | 46.51 | 80.32 |
| Node 21 | Mass flow rate | $\varepsilon_1(10^{-2})$ | 3.96 | 3.99 | 3.99 | 3.99 |
| | | $\varepsilon_2(10^{-5})$ | 2.22 | 3.26 | 4.48 | 8.33 |
| Bus 11 | $e$ | $\varepsilon_1(10^{-3})$ | 2.67 | 1.89 | 1.39 | 0.80 |
| | | $\varepsilon_2(10^{-6})$ | 1.31 | 1.50 | 1.71 | 1.72 |
| | $f$ | $\varepsilon_1(10^{-3})$ | 2.24 | 2.07 | 1.70 | 1.08 |
| | | $\varepsilon_2(10^{-6})$ | 1.13 | 1.72 | 2.31 | 2.33 |

The filter coefficients and total variances are presented in Tab. III. It is can be seen that the filter coefficients are still small than 1 even the error deviations exist. With the deviations increasing, the total variances increase also, while the filter coefficients show a different condition. For the node pressure, the filter coefficients increase because the estimating errors enlarge due to the bigger deviations. For the mass flow rate, the filter coefficients are almost same under different deviations, since the changing degree of estimating errors is similar to the measurement errors. For the bus voltage, the filter coefficients decrease due to the big deviations of measurement errors. It is can be concluded that the node pressure estimating results are the most sensitive to the measurement error deviations.

*C. Non-Gaussian Noises*

The measurement errors always satisfy the Cauchy distribution or Laplace distribution [17], [19]. The errors satisfying Cauchy and Laplace distributions are added to the true values of voltages and gas pipeline states respectively, which are taken as measurements, to test the performances of the proposed DSE method.

The indexes are shown in Tab. IV. It is can be seen that the filter coefficients are smaller than 1, which means that the proposed DSE method is effective under the Cauchy and Laplace distributions. Compared with Gaussian noises, the total variances of the Laplace distribution are almost the same, while the values of the Cauchy distribution are much larger, indicating that the estimating performances are much sensitive to the Cauchy distribution than the Laplace. The reason is that the random numbers deviating seriously from the location parameter in the Cauchy distribution are more than the ones in the Laplace distribution, which may decrease the estimating accuracy of the DSE.

The simulations are carried out on the computer with 1.60GHz CPU, and the average time consumption of one step is 7.928ms, which can completely meet the real-time requirements in practical applications.

VI. CONCLUSION

A DSE method for IGESs is proposed in the paper, which is applied on an IGES with 30-node pipeline network and IEEE 39-bus system coupled by two GTUs, under several measurement conditions. The filter coefficient and total variance are used to evaluate the performances of the DSE method. The results show that all of the filter coefficients are smaller than 1 under the white Gaussian noises condition, meaning that the DSE is effective. Furthermore, the method is studied under the non-zero mean and non-Gaussian noises conditions. The results show that the node pressure is more sensitive to the non-zero mean noises, and the estimating performances decrease further with Cauchy distribution than Laplace.

TABLE IV. THE INDEXES UNDER CAUCHY AND LAPLACE DISTRIBUTION

| | Cauchy distribution | | | | Laplace distribution | | | |
|---|---|---|---|---|---|---|---|---|
| | $e$ | | $f$ | | $e$ | | $f$ | |
| Bus | $\varepsilon_1$ | $\varepsilon_2(10^{-3})$ | $\varepsilon_1$ | $\varepsilon_2(10^{-3})$ | $\varepsilon_1$ | $\varepsilon_2(10^{-6})$ | $\varepsilon_1$ | $\varepsilon_2(10^{-6})$ |
| 1 | 0.1846 | 0.7271 | 0.0384 | 0.2227 | 0.0049 | 0.4894 | 0.0036 | 0.4856 |
| 2 | 0.0010 | 0.7217 | 0.0012 | 0.2206 | 0.0051 | 0.4252 | 0.0052 | 0.4834 |
| 4 | 0.0163 | 0.7196 | 0.1568 | 0.2207 | 0.0039 | 0.4345 | 0.0039 | 0.4854 |
| 9 | 0.0204 | 0.7254 | 0.016 | 0.2229 | 0.0049 | 0.4591 | 0.0043 | 0.4852 |
| 12 | 0.2164 | 0.7152 | 0.0025 | 0.2206 | 0.0068 | 0.4610 | 0.0057 | 0.4968 |
| 15 | 0.0101 | 0.7254 | 0.0711 | 0.2205 | 0.0043 | 0.4291 | 0.0047 | 0.4717 |
| 18 | 0.1860 | 0.7236 | 0.0100 | 0.2203 | 0.0059 | 0.4187 | 0.0033 | 0.4855 |
| 22 | 0.5017 | 0.7345 | 0.0047 | 0.2202 | 0.0047 | 0.4462 | 0.0071 | 0.4771 |
| 24 | 0.0706 | 0.7274 | 0.0783 | 0.2202 | 0.0052 | 0.4267 | 0.0030 | 0.4696 |
| 27 | 0.0114 | 0.7248 | 0.0295 | 0.2206 | 0.0038 | 0.4189 | 0.0055 | 0.4843 |
| 30 | 0.0123 | 0.7181 | 0.0702 | 0.2192 | 0.0049 | 0.4335 | 0.0060 | 0.4943 |
| 31 | 0.0037 | 0.7163 | 0.0085 | 0.2191 | 0.0042 | 0.4226 | 0.0057 | 0.5052 |
| 32 | 0.1235 | 0.7174 | 0.0021 | 0.2204 | 0.0054 | 0.4584 | 0.0050 | 0.5090 |
| 35 | 0.0071 | 0.7333 | 0.0488 | 0.2196 | 0.0040 | 0.4578 | 0.0052 | 0.4872 |
| 36 | 0.0002 | 0.7519 | 0.0017 | 0.2187 | 0.0047 | 0.4715 | 0.0056 | 0.4894 |
| 38 | 0.1093 | 0.0038 | 0.0528 | 0.2174 | 0.0046 | 0.4733 | 0.0055 | 0.5660 |
| | $p$ | | $\dot{m}$ | | $p$ | | $\dot{m}$ | |
| Node | $\varepsilon_1(10^{-3})$ | $\varepsilon_2(10^{-5})$ | $\varepsilon_1$ | $\varepsilon_2$ | $\varepsilon_1(10^{-4})$ | $\varepsilon_2(10^{-8})$ | $\varepsilon_1$ | $\varepsilon_2(10^{-4})$ |
| 3 | 0.0001 | 0.0015 | 0.2241 | 0.0024 | 0.0005 | 0.0005 | 0.2243 | 0.3063 |
| 5 | 0.0029 | 0.0049 | 0.2243 | 0.0129 | 0.0029 | 0.0026 | 0.2242 | 0.3638 |
| 7 | 0.0568 | 0.0168 | 0.2244 | 0.2025 | 0.0090 | 0.0094 | 0.2239 | 0.3993 |
| 8 | 0.0001 | 0.0004 | 0.2244 | 0.0170 | 0.0004 | 0.0003 | 0.2226 | 0.2152 |
| 10 | 0.0001 | 0.1162 | 0.0532 | 0.0003 | 0.0188 | 0.0136 | 0.0524 | 0.0429 |
| 11 | 0.0405 | 0.0126 | 0.0533 | 0.0287 | 0.0114 | 0.0135 | 0.0527 | 0.0539 |
| 13 | 0.0525 | 0.0316 | 0.1406 | 0.0013 | 0.0201 | 0.0163 | 0.1402 | 0.1417 |
| 15 | 0.0103 | 0.0043 | 0.1386 | 0.0005 | 0.0278 | 0.0269 | 0.1388 | 0.1173 |
| 16 | 0.0054 | 0.0079 | 0.0533 | 0.0004 | 0.6771 | 0.6299 | 0.2323 | 0.1260 |
| 17 | 0.0124 | 0.0157 | 0.0532 | 0.0079 | 0.2417 | 0.1660 | 0.0837 | 0.1026 |
| 20 | 0.0007 | 0.0146 | 0.1406 | 0.0009 | 0.0419 | 0.0298 | 0.1403 | 0.1701 |
| 22 | 0.0078 | 0.0066 | 0.1371 | 0.0015 | 0.0226 | 0.0192 | 0.1401 | 0.1748 |
| 23 | 0.0037 | 0.0326 | 0.2244 | 0.0013 | 0.0217 | 0.0209 | 0.2238 | 0.2726 |
| 24 | 0.0321 | 0.0304 | 0.0533 | 0.0002 | 0.0226 | 0.0255 | 0.0532 | 0.0587 |
| 27 | 0.0632 | 0.8513 | 0.0533 | 0.5381 | 0.0190 | 0.0246 | 0.0507 | 0.0487 |
| 29 | 0.0701 | 0.0071 | 0.2241 | 0.0024 | 0.0337 | 0.0327 | 0.0516 | 0.0527 |

As a complete new research field, further works of the DSE for IGESs can be carried out in the following areas: the accurate calculating method of predicting errors; the robust DSE algorithm against bad data, colored noises and model parameter errors; the distributed DSE method for IGESs.